\documentclass[aps,floatfix,onecolumn,a4paper,noshowpacs, nofootinbib,superscriptaddress,11pt]{revtex4}
\usepackage{amssymb,amsmath,amsfonts}
\usepackage{subfigure}
\usepackage[dvips]{graphics,color}
\usepackage{epsfig}\usepackage{float}

\usepackage{hyperref,color,xcolor}
\hypersetup{colorlinks=true, linkcolor=blue, citecolor=blue, urlcolor=blue}

\hypersetup{hidelinks,colorlinks=true,breaklinks=true,urlcolor= blue}
\hypersetup{%
    colorlinks = true,
    linkcolor  = blue,
    citecolor = blue,
  }
\usepackage{booktabs} 
\usepackage{array}
\newcommand{\hko}{\hookrightarrow}

\newcommand{\n}{\nonumber\\}

\newcommand{\bu}{\bullet}

\newcommand{\cv}{\circ}

\newcommand{\bec}{\begin{center}}
\newcommand{\eec}{\end{center}}

\newcommand{\bea}{\begin{array}}
\newcommand{\ear}{\end{array}}

\newcommand{\bfr}{\begin{flushright}}

\newcommand{\efr}{\end{flushright}}
\newcommand{\noi}{\noindent}\newcommand{\Ra}{\rightarrow}

\newcommand{\cl}{{\mt{C}}\ell}

\newcommand{\RR}{\mathbb{R}}\newcommand{\op}{\oplus}

\newcommand{\la}{\Lambda}
\newcommand{\bege}{\begin{equation}}
\newcommand{\enge}{\end{equation}}

\newcommand{\beq}{\begin{eqnarray}}\newcommand{\benu}{\begin{enumerate}}\newcommand{\enu}{\end{enumerate}}
\newcommand{\eeq}{\end{eqnarray}}
\newcommand{\mt}{\mathcal}

\newcommand{\CC}{\mathbb{C}}

\newcommand{\OO}{\mathbb{O}}

\newcommand{\mk}{\mathfrak}

\newcommand{\bx}{\begin{pmatrix}}
\newcommand{\ex}{\end{pmatrix}}

\usepackage{indentfirst}  
\begin{document}
\title{Exceptional $\mathfrak{g}_2$ deformations and gauge symmetries}
\author{G. Karapetyan}
\email{gayane.karapetyan@ufabc.edu.br}
\affiliation{Federal University of ABC, Center of Mathematics, Santo Andr\'e, S\~ao Paulo, 09580-210, Brazil}

\begin{abstract} Deformed $\mathfrak{g}_2$ exceptional applications are introduced via the Clifford algebra-parametrized formalism. Using the products between multivectors 
of $\cl_{0,7}$, the Clifford algebra over the metric vector space $\RR^{0,7}$,   and octonions, resulting in an octonion, we generalize the exceptional Lie algebra $\mathfrak{g}_2$ applications, also associated with the transformation rules for bosonic and fermionic fields 
on the 7-sphere $S^7$. The emergence of $SU(3)$-like subalgebras within the exceptional Lie algebra $\mathfrak{g}_2$ provides an algebraic framework reminiscent of the $SU(3)$ gauge symmetry of QCD.
\end{abstract}
\maketitle

\section{Introduction}

The $X$-product was originally introduced to correctly define 
the transformation rules for bosonic and fermionic fields 
on the 7-sphere $S^7$ \cite{ced}. This product is 
closely related to the parallel transport of sections of the tangent bundle, at $X\in S^7$, i.e., $X\in\OO$ is an unitary element of the octonion algebra $\OO$. 
The $X$-product is also shown to be twice 
the parallelizing torsion. The $X$-product  has also been used to obtain triality maps and $G_2$ applications \cite{dix,beng},
and it leads naturally to remarkable geometric and topological properties, for instance the Hopf fibrations  \cite{daRocha:2008we,daRocha:2009gb}, and twistor formalism in ten dimensions \cite{mart}.
Other applications of the formalism presented in Ref. \cite{eu1} can be viewed in \cite{daro1}. 

Although quantum chromodynamics (QCD) is formulated as an associative Yang--Mills theory with gauge group $SU(3)$, it is well known that $SU(3)$ can be embedded into the exceptional Lie group $G_2$, the automorphism group of the octonion algebra. In particular, $SU(3)$ arises as the stabilizer of a unit imaginary octonion, making color symmetry naturally interpretable as a residual symmetry of an underlying octonionic structure. This observation is further supported by lattice studies of $G_2$ Yang--Mills theory, which exhibit confinement, a rich glueball spectrum, and nonperturbative dynamics closely analogous to those of QCD, despite the trivial center of $G_2$ \cite{Masi:2021cgm,Holland:2003jy,Wellegehausen:2011sc}. Such results indicate that several key features of strong interactions are not uniquely tied to $SU(3)$ gauge symmetry, but may instead reflect more general properties of non-Abelian gauge theories \cite{Bernardini:2018uuy,Ferreira:2019inu,Correa:2015lla}. Motivated by these insights, the octonionic $X$-product and its generalizations provide a natural framework for exploring deformations of $G_2$-based algebraic structures. Although the present work does not aim to construct QCD directly, it offers an exceptional geometric setting in which QCD-like gauge symmetries emerge from deformations of non-associative octonionic products, suggesting a possible higher-dimensional or pre-associative origin of color dynamics.
 
The importance of octonions in the search for unification arises from their unique algebraic properties. As the largest normed division algebra, octonions are non-associative yet alternative, and their structure is intimately linked to exceptional Lie groups and higher-dimensional spinors \cite{daRocha:2007sd}. These features make them a natural framework for describing the symmetries of extended objects, triality, and exceptional geometries, all of which play crucial roles in modern attempts at unifying the fundamental interactions \cite{Yanes:2018krn,Vaz:2016qyw}.

One of the most striking applications of octonions in theoretical physics is their role in higher-dimensional supersymmetry. By extending standard division-algebra-valued superalgebras to octonionic entries, one can construct, in eleven dimensions ($D=11$), an octonionic generalized Poincar\'e superalgebra, commonly referred to as the octonionic $M$-algebra. This algebra provides an elegant and unified description of the symmetries underlying $M$-theory, in particular capturing the dynamics of extended supersymmetric objects. Within this framework, the octonionic super-2-brane (M2-brane) and super-5-brane (M5-brane) sectors are shown to be equivalent, a remarkable feature that is deeply connected to the properties of the octonionic product and the geometry of the 7-sphere $S^7$ \cite{Ivanova:1993nu,Rooman:1983es,daRocha:2021xwq}. The $S^7$ structure, parallelizable and intimately related to octonions, provides a natural geometric arena in which these extended supersymmetries and dualities are realized.

Beyond $M$-theory, octonions have far-reaching applications across theoretical physics. Their automorphism group, $G_2$, along with other exceptional groups derived from octonionic constructions, plays a central role in grand unified theories, string compactifications, and duality symmetries. Octonionic triality underlies the equivalence between vector, spinor, and conjugate spinor representations in $\mathrm{Spin}(8)$, which is crucial for the construction of higher-dimensional spinors and supersymmetric multiplets 
\cite{Gunaydin:2020ric,Borsten:2008wd,Krausshar:2020bse,loun}. Furthermore, octonion-valued gauge theories have been explored, revealing novel instanton solutions and self-duality conditions in higher dimensions. The exceptional geometric structures encoded by octonions, including their role in Hopf fibrations and twistor spaces, provide both a geometric and algebraic foundation for the study of extended supersymmetry and exceptional compactifications \cite{Gonzales:2009ye,Kuznetsova:2006ws,Carrion:2002ri,Lukierski:2002ux}.

In addition to their formal algebraic significance, octonions serve as a bridge between abstract mathematics and physical phenomena. Their non-associativity manifests as torsion in $S^7$ compactifications, while their normed property ensures the consistency of higher-dimensional spinor bilinears \cite{Ablamowicz:2014rpa,Bonora:2014dfa,Lopes:2018cvu,Bonora:2015ppa,Goncalves:2023pty,CoronadoVillalobos:2015mns}. Triality symmetries of octonions reflect the duality relations between extended objects in $D=11$ spacetime, providing deep insight into the structure of $M$-theory. As a result, octonions are not merely a mathematical curiosity but a fundamental tool in the ongoing search for unified theories \cite{daRocha:2017cxu}, offering a rich algebraic and geometric framework that continues to inspire developments in supergravity, string theory, and beyond \cite{Gunaydin:1974fb,Abdalla:2009pg,daRocha:2004uf}.

Applications of octonions continue to expand, encompassing the classification of extended supersymmetries, constructions of octonionic versions of the Standard Model, explorations of quantum information in exceptional Jordan algebras, and studies of exotic compactifications in string theory 
\cite{Meert:2020sqv,Casadio:2017sze,daRocha:2020gee}. Their exceptional nature ensures that each new application illuminates both the algebraic and geometric foundations of fundamental physics, making octonions a central object in the ongoing quest to understand the underlying symmetries of the universe \cite{baez,top,carr,Aizawa:2017nue,okubo}.

The main aim of this paper is to introduce some deformed $\mathfrak{g}_2$ exceptional applications from the Clifford algebra-parametrized formalism. 
 This paper is organized as follows: Section II is devoted to presenting exterior and Clifford algebras, and  
in Section III, we review the 
fundamental properties of the octonionic algebra $\OO$, defined in terms of the Clifford algebra $\cl_{0,7}$ and its associated Clifford product
that defines the octonionic product \cite{loun}. The octonionic product is chosen to be defined in terms of the Clifford product 
so that the arena of the whole formalism is the Clifford algebra. In Section IV, the $X$-product and the $XY$-product are generalized, encompassing 
octonionic-generalized products between arbitrary multivectors of $\cl_{0,7}$ and octonions.
After introducing products 
between Clifford multivectors and octonions, that results in an octonion, we illustrate the use of the formalism, giving 
some examples of useful computations. 
In Section V, we analyze how the exceptional Lie algebra $\mathfrak{g}_2$ admits $\mathfrak{su}(3)$-like subalgebras as stabilizers of distinguished octonionic directions, and how Clifford-parametrized deformations interpolate between inequivalent embeddings, yielding residual $SU(3)$-like symmetries. Section VI is devoted to introducing some deformed $\mathfrak{g}_2$ exceptional applications from the Clifford algebra-parametrized formalism.

\section{Preliminaries}
Let $V$ be a finite-dimensional real vector space of dimension $n$, and let $V^*$ denote its dual space. Consider the tensor algebra $T(V) = \bigoplus_{i=0}^\infty T^i(V)$, from which we focus on the subspace of multivectors $\Lambda(V) = \bigoplus_{k=0}^n \Lambda^k(V)$, where $\Lambda^k(V)$ denotes the space of antisymmetric $k$-tensors, naturally isomorphic to the space of $k$-forms.  
For a multivector $\psi \in \Lambda(V)$, the reversion (an algebra antiautomorphism) is defined as $\tilde{\psi} = (-1)^{[k/2]} \psi$ for $\psi \in \Lambda^k(V)$, where $[k]$ is the integer part of $k$. The main automorphism (or graded involution) is given by $\hat{\psi} = (-1)^k \psi$, and the conjugation is the composition of reversion followed by the main automorphism.  
If $V$ is equipped with a non-degenerate symmetric bilinear form $g: V^* \times V^* \to \mathbb{R}$, it can be extended naturally to $\Lambda(V)$ by $g(\psi, \phi) = \det(g(u^i, v^j))$ if $\psi = u^1 \wedge \cdots \wedge u^k$ and $\phi = v^1 \wedge \cdots \wedge v^k$, and $g(\psi, \phi) = 0$ if the degrees of $\psi$ and $\phi$ differ.  
The projection of a multivector $\psi = \psi_0 + \psi_1 + \cdots + \psi_n$, with $\psi_k \in \Lambda^k(V)$, onto its $p$-vector part is denoted by $\langle \psi \rangle_p = \psi_p$.  
Given $\psi, \phi, \xi \in \Lambda(V)$, the left contraction is defined implicitly through $g(\psi \lrcorner \phi, \xi) = g(\phi, \tilde{\psi} \wedge \xi)$ \cite{Vaz:2016qyw}. In particular, for scalars $a \in \mathbb{R}$, ${ v} \lrcorner a = 0$. The contraction obeys the Leibniz rule ${ v} \lrcorner (\psi \wedge \phi) = ({ v} \lrcorner \psi) \wedge \phi + \hat{\psi} \wedge ({ v} \lrcorner \phi)$ for ${ v} \in V$.  
Similarly, the right contraction is defined by $g(\psi \llcorner \phi, \xi) = g(\phi, \psi \wedge \tilde{\xi})$, with Leibniz rule $(\psi \wedge \phi) \llcorner { v} = \psi \wedge (\phi \llcorner { v}) + (\psi \llcorner { v}) \wedge \hat{\phi}$. The left and right contractions are related by ${ v} \lrcorner \psi = - \hat{\psi} \llcorner { v}$.  
The Clifford product between ${ w} \in V$ and $\psi \in \Lambda(V)$ is then ${ w} \psi = { w} \wedge \psi + { w} \lrcorner \psi$.  
The Grassmann algebra $(\Lambda(V), g)$ equipped with the Clifford product defines the Clifford algebra $\mathrm{Cl}(V,g) \simeq \mathrm{Cl}_{p,q}$, with $V \simeq \mathbb{R}^{p,q}$ and $p + q = n$.  
Throughout, $\mathbb{R}, \mathbb{C}, \mathbb{H}$ denote the real, complex, and quaternionic number fields, respectively.

\section{Octonions}
The octonion algebra $\mathbb{O}$ is defined as the paravector space $\mathbb{R} \oplus \mathbb{R}^{0,7}$ equipped with the octonionic product $\circ : (\mathbb{R} \oplus \mathbb{R}^{0,7}) \times (\mathbb{R} \oplus \mathbb{R}^{0,7}) \to \mathbb{R} \oplus \mathbb{R}^{0,7}$. The identity $e_0 = 1 \in \mathbb{R}$ together with an orthonormal basis $\{e_a\}_{a=1}^7$ in the paravector space $\mathbb{R} \oplus \mathbb{R}^{0,7} \subset \mathrm{Cl}_{0,7}$ generates the octonions.  
It is well known that the octonionic product can be constructed using the Clifford algebra $\mathrm{Cl}_{0,7}$ as
$A \circ B = \langle AB(1 - \psi) \rangle_{0 \oplus 1}$, for $A, B \in \mathbb{R} \oplus \mathbb{R}^{0,7}$,  
where $\psi = e_1 e_2 e_4 + e_2 e_3 e_5 + e_3 e_4 e_6 + e_4 e_5 e_7 + e_5 e_6 e_1 + e_6 e_7 e_2 + e_7 e_1 e_3 \in \Lambda^3(\mathbb{R}^{0,7}) \subset \mathrm{Cl}_{0,7}$, and the juxtaposition denotes the Clifford product.  

The idea of introducing the octonionic product from the Clifford product is to present our formalism using Clifford algebras exclusively. Since $\mathbb{O}$ is isomorphic to $\mathbb{R} \oplus \mathbb{R}^{0,7}$ as a vector space, the octonionic product takes two arbitrary elements of the paravector space and produces another element in the same space. By considering octonions inside the Clifford algebra, one can go beyond the paravector space and exploit the entire Clifford algebra, which allows generalizations such as the $X$- and $XY$-products.  

From the above definition, one can verify the usual multiplication rules between basis elements:  
$e_a \circ e_b = c_{ab}^c e_c - \delta_{ab}$ for $a,b,c=1,\dots,7$,  
where $c_{ab}^c = 1$ for the cyclic permutations $(abc) = (124), (235), (346), (457), (561), (672), (713)$. Explicitly, the multiplication table is given by  
\begin{table}[h!]
\centering
\renewcommand{\arraystretch}{1.3} 
\setlength{\tabcolsep}{6pt} 
\begin{tabular}{!{\vrule width 1pt}r!{\vrule width 1pt}*{7}{c!{\vrule width 1pt}}}
\specialrule{1pt}{0pt}{0pt}  
 1 & $e_1$ & $e_2$ & $e_3$ & $e_4$ & $e_5$ & $e_6$ & $e_7$ \\
\specialrule{1pt}{0pt}{0pt}  
$e_1$ & $-1$ & $e_4$ & $e_7$ & $-e_2$ & $e_6$ & $-e_5$ & $-e_3$ \\
\hline
$e_2$ & $-e_4$ & $-1$ & $e_5$ & $e_1$ & $-e_3$ & $e_7$ & $-e_6$ \\
\hline
$e_3$ & $-e_7$ & $-e_5$ & $-1$ & $e_6$ & $e_2$ & $-e_4$ & $e_1$ \\
\hline
$e_4$ & $e_2$ & $-e_1$ & $-e_6$ & $-1$ & $e_7$ & $e_3$ & $-e_5$ \\
\hline
$e_5$ & $-e_6$ & $e_3$ & $-e_2$ & $-e_7$ & $-1$ & $e_1$ & $e_4$ \\
\hline
$e_6$ & $e_5$ & $-e_7$ & $e_4$ & $-e_3$ & $-e_1$ & $-1$ & $e_2$ \\
\hline
$e_7$ & $e_3$ & $e_6$ & $-e_1$ & $e_5$ & $-e_4$ & $-e_2$ & $-1$ \\
\specialrule{1pt}{0pt}{0pt}  
\end{tabular}
\caption{Octonionic multiplication table. All the relations above can be expressed as $e_a\cv e_{a+1} = e_{a+3\mod 7}$.}
\label{tabela}
\end{table}

Since the underlying vector space of $\mathbb{O}$ is $\mathbb{R} \oplus \mathbb{R}^{0,7} \subset \mathrm{Cl}_{0,7}$, the Clifford conjugation of $X = x^0 + x^a e_a \in \mathbb{O}$ is $\bar{X} = x^0 - x^a e_a$.

\section{The $u$-product and generalizations}
Consider $u\in\cl_{0,7}\equiv\cl(\RR^{0,7},g)$. 
Given fixed but arbitrary  $X,Y \in\RR\op\RR^{0,7}$ such that $X\bar{X} = \bar{X} X = 1 = \bar{Y}Y = Y\bar{Y}$ ($X,Y\in S^7$),
 the $X$-product is defined \cite{ced,mart,dix} by 
\bege
A\circ_X B:= (A\cv X)\circ(\bar{X} \cv B),
\enge\noi 
whereas the $XY$-product is defined as:
\bege\label{xy}
A\circ_{X,Y} B:= (A\cv X)\cv(\bar{Y}\cv B).
\enge\noi In particular, the $(1,X)$-product is given by 
\bege\label{pro3}
A\circ_{1,X} B:= A\cv(\bar{X}\cv B).
\enge\noi  $X$ is the unit of $(1,X)$-product above, since $A\circ_{1,X} X = X\circ_{1,X} A = A$ \cite{ced,dix}.

For  homogeneous multivectors  $u = u_1\ldots u_k\in \la^k(\RR^{0,7}) \hko \cl_{0,7}$, where $\{u_p\}_{p=1}^k \subset \RR^{0,7}$ ($k=1,\ldots,7)$ and 
$A\in\RR\op\RR^{0,7}$, one defines the product $\bu_\llcorner$ as \cite{eu1,daRocha:2012tw}
\beq\label{def1}
\bu_\llcorner: (\RR\op\RR^{0,7})\times\la^k(\RR^{0,7}) &\Ra& \RR\op\RR^{0,7}\nonumber\\
(A,u) &\mapsto& A\bu_\llcorner u = ((\cdots(A\cv u_1)\cv u_2)\cv\cdots)\cv u_{k-1})\cv u_k.\label{def2}
\eeq\noi The symbol $\bu_\llcorner$ indicates that the octonion element $A$  appears in the left entry of the product.

One also defines the product $\bu_\lrcorner$ as 
\beq\bu_\lrcorner: \la^k(\RR^{0,7}) \times (\RR\op\RR^{0,7}) &\Ra& \RR\op\RR^{0,7}\nonumber\\
(u,A)&\mapsto& u\bu_\lrcorner A =  u_1\cv(\cdots\cv(u_{k-1}\cv(u_k\cv A))\cdots),\label{d3}
\eeq\noi with the symbol $\bu_\lrcorner$ remembers us the element $A$ enters in the \emph{right} entry in the product in Eq. (\ref{d3}).

\medbreak
{\bf Remark 1}: It is clear that by extending the product in Eq. (\ref{def2}) to the scalars --- elements of $\la^0(\RR^{0,7})$ --- we have that $A \bu_\llcorner a = aA$, 
which denotes the trivial multiplication by scalars, where $a\in \RR = \la^0(\RR^{0,7})$.
Therefore, one extends by linearity the product $\bu_\llcorner$ to the entire exterior algebra $\la(\RR^{0,7}) = \oplus_{a=0}^7\la^a(\RR^{0,7})$:
\beq\label{def10}
\dot\bu_\llcorner: (\RR\op\RR^{0,7})\times\la(\RR^{0,7}) &\Ra& \RR\op\RR^{0,7}\\
\dot\bu_\lrcorner: \la(\RR^{0,7}) \times (\RR\op\RR^{0,7}) &\Ra& \RR\op\RR^{0,7}\label{def11}
\eeq
 \medbreak 
By abuse of notation we shall use hereon the symbol $\bu$ uniquely to denote both products $\dot\bu_\llcorner$ and $\dot\bu_\lrcorner$, 
in Eq. (\ref{def10}) and Eq. (\ref{def11}),
 and each one of the above-mentioned products is to be clearly implicit, as there exists an octonion in the left \emph{or} right entry 
of the product $\bu$.

Now, given $u\in\la(\RR^{0,7})$, the $u$-product can be defined as
\beq
\circ_u: (\RR\op\RR^{0,7})\times(\RR\op\RR^{0,7})&\Ra&(\RR\op\RR^{0,7})\n
(A,B)&\mapsto& A\circ_u B:= (A\bu u)\cv(u^{-1}\bu B).
\eeq\noi For invertible elements $u\in\cl_{0,7}$, it follows that 
\bege\label{2222}
A\cv_u B = (A\cv(B\bu u))\bu u^{-1} = u\bu((u^{-1}\bu A)\cv B).
\enge
 When  $u$ is a paravector it is clear that the $u$-product is equivalent to the $X$-product.

Similarly to the $XY$-product and the ($1,X$)-product, respectively at Eqs. (\ref{xy}) and (\ref{pro3}), 
it is also possible to define the $(1,u)$-product, as 
\beq\label{jiri}
\cv_{1,u}:(\RR\op\RR^{0,7})\times(\RR\op\RR^{0,7}) &\Ra& \RR\op\RR^{0,7}\n
(A,B)&\mapsto& A\circ_{1,u} B:= A\cv(u^{-1}\bu B).
\eeq \noi
Finally, Eq. (\ref{xy}) can be generalized, given fixed $u,v\in\cl_{0,7}$, as:
\beq\label{uvb}
\cv_{u,v}:(\RR\op\RR^{0,7})\times(\RR\op\RR^{0,7}) &\Ra& \RR\op\RR^{0,7}\n
(A,B)&\mapsto& A\circ_{u,v} B:= (A\bu u)\cv(v^{-1}\bu B).
\eeq

\section{SU(3)-like subalgebras and residual symmetry.}
It is well known that the exceptional Lie algebra $\mathfrak{g}_2=\mathrm{der}(\OO)$ admits
$\mathfrak{su}(3)$ as a maximal subalgebra, realized as the stabilizer of a fixed unit imaginary
octonion. Explicitly, fixing $n\in\mathrm{Im}\,\OO$ with $n^2=-1$, one has
\begin{equation}
\mathfrak{su}(3)\;\simeq\;\{\,D\in\mathfrak{g}_2\mid D(n)=0\,\},
\end{equation}
and the induced decomposition $\OO=\RR\oplus\RR n\oplus n^\perp$, with $n^\perp\simeq\CC^3$ via
right multiplication by $n$, yields the Lie algebra splitting
\begin{equation}
\mathfrak{g}_2 \;\simeq\; \mathfrak{su}(3)\;\oplus\;\mathbf{3}\;\oplus\;\bar{\mathbf{3}}.
\end{equation}
Here $\mathfrak{su}(3)$ acts faithfully on the complex triplet sector while leaving $n$
invariant, and the $\mathbf{3}\oplus\bar{\mathbf{3}}$ components are generated by derivations
that do not preserve the chosen imaginary direction.

Within the present Clifford-parametrized framework, the deformed octonionic products
$\circ_u$, $\circ_{1,u}$, and $\circ_{u,v}$ induce corresponding deformations of the derivation
algebra, $\mathfrak{g}_2\to(\mathfrak{g}_2)_u$, as established in Lemmas~1 and~2. For generic
$u\in\cl_{0,7}$, the stabilizer of a distinguished imaginary direction in $\OO_u := (\OO, \circ_u)$ is no longer
strictly $\mathfrak{su}(3)$, but rather a conjugate or deformed $\mathfrak{su}(3)$-like
subalgebra inside $(\mathfrak{g}_2)_u$, generated by derivations preserving the complex
structure induced by $u$. In this sense, the $u$-deformation interpolates between inequivalent
embeddings of $\mathfrak{su}(3)$ inside $\mathfrak{g}_2$, providing a controlled algebraic
mechanism by which $SU(3)$-like symmetries arise as residual symmetries of exceptional,
non-associative structures.

\section{Exceptional $\mathfrak{g}_2$ deformed applications}
Given $u\in\cl_{0,7}$, by Eq. (\ref{2222}) we can state that
\beq
A\cv_u B &=& (A\bu u)\cv(u^{-1}\bu B) = ((A\bu u^{2/3})\bu(u^{1/3})\cv
 (u^{-1/3}\bu(u^{-2/3}\bu B))\n
&=& u^{1/3}\bu[(u^{-1/3}\bu(A\bu u^{2/3}))\cv(u^{-2/3}\bu B)], 
 \quad\text{using Eq. (\ref{2222})}
\n 
&=& u^{1/3}\bu[(u^{-1/3}\bu (A\bu u^{1/3}\bu u^{1/3}))\cv
(u^{-1/3}\bu u^{-1/3}\bu B)],
\n 
&=& u^{1/3}\bu([(u^{-1/3}\bu (A\bu u^{1/3}))\cv
(u^{-1/3}\bu B\bu u^{1/3})]\bu u^{-1/3})\label{31}\\
&=& u^{1/3}\bu [(u^{-1/3}\bu (A\bu u^{1/3}))\cv
(u^{-1/3}\bu B\bu u^{1/3})]\bu u^{-1/3} 
 \eeq
It is immediate to see that in the particular case where $u = \pm 1$, it follows that
\beq
AB &=& u^{1/3}\bu [(u^{-1/3}\bu (A\bu u^{1/3}))\cv
(u^{-1/3}\bu B\bu u^{1/3})]\bu u^{-1/3} 
\eeq and consequently
\beq\label{auto}
u^{-1/3}\bu (AB) \bu u^{1/3} &=&   [(u^{-1/3}\bu A\bu u^{1/3})\cv
(u^{-1/3}\bu B\bu u^{1/3})] 
\eeq i.e., the application 
\beq
f_u:&& \OO \rightarrow \OO\nonumber\\
&&A\mapsto f_u(A) := u^{-1/3}\bu A \bu u^{1/3}
\eeq\noi is indeed in the group $G_2$ of the automorphisms of the algebra $\OO$. This construction yields an algebra isomorphic to the standard octonions, $\OO$, and it admits its own group of automorphisms $G_2{}_u$. Consequently, the automorphisms of the twisted algebra $\OO_u$ are conjugate to the standard $G_2$ automorphisms via $f_u$, i.e.,
\beq
G_2{}_u = f_u^{-1} \circ G_2 \circ f_u.
\eeq

At the Lie algebra level, the corresponding $u$-twisted Lie algebra is naturally defined as the derivation algebra of $\OO_u$:
\beq
(\mathfrak{g}_2){}_u := \mathrm{der}(\OO_u) = \{ D \in \mathrm{End}(\OO) \mid D(A \circ_u B) = D(A) \circ_u B + A \circ_u D(B), \;\forall A,B\in \OO \}.
\eeq
Here, $\mathrm{der}(\OO_u)$ consists of all linear maps that satisfy the Leibniz rule with respect to the twisted multiplication $\circ_u$. Because $\OO_u$ is isomorphic to $\OO$ via $f_u$, every derivation $D \in \mathfrak{g}_2{}_u$ is conjugate to a standard derivation in $\mathfrak{g}_2 = \mathrm{der}(\OO)$:
\beq
D \in (\mathfrak{g}_2){}_u \quad \iff \quad f_u \circ D \circ f_u^{-1} \in \mathfrak{g}_2.
\eeq
Thus, the algebra $\mathfrak{g}_2{}_u$ captures all infinitesimal symmetries of the twisted algebra $\OO_u$ in the same way that $(\mathfrak{g})_2$ does for $\OO$. It forms the Lie algebra of the twisted automorphism group $G_2{}_u$. This construction provides a natural framework to study the interplay between octonionic multiplication, conjugation by powers of $u$, and the exceptional Lie group $G_2$ in both the algebra and its derivations.

Given $A,B\in \OO$, recall that  $A\circ_u B:= (A\bu u)\cv(u^{-1}\bu B)$. We enunciate 
\medbreak {\bf Lemma 1}: \emph{If} $h\in\mk{g}_2 = \mk{der}(\OO)$, \emph{for arbitrary invertible homogeneous elements}  $u\in \cl_{0,7}$ \emph{such that} $u\neq \pm 1$, \emph{then}
 $$u^{1/3} \bu [h(u^{-1/3}\bu (\;\cdot\;) \bu u^{1/3})]\bu u^{-1/3}\in{(\mk{g}_2)}_u \equiv \mk{der}(\OO_u).$$
 \medbreak
 \emph{By defining the application}
  \beq M_{h,u}: \OO &\rightarrow& \OO\n
A&\mapsto& M_{h,u}(A):= u^{1/3} \bu  [h(u^{-1/3}\bu (A) \bu u^{1/3})]\bu u^{-1/3},\nonumber\eeq
 \emph{it satisfies} 
 \beq{M_{h,u} (A\circ_u B) = M_{h,u}(A)\circ_u B + A \circ_u M_{h,u}(B)},\eeq \emph{i.e}, 
 $M_{h,u}$ \emph{is a derivation of} $\OO_u$.\\
 \medbreak
 {\bf Proof}: Take $A,B\in\OO$ and an invertible homogeneous element $u\in\cl_{0,7}$, then
 \beq
M_{h,u} (A\circ_u B)&=& u^{1/3} \bu [h(u^{-1/3}\bu (A\circ_u B) \bu u^{1/3})]\bu u^{-1/3}\nonumber\\
&=& u^{1/3} \bu [h(u^{-1/3}\bu A \bu u^{1/3})\circ (u^{-1/3}\bu B \bu u^{1/3})]\bu u^{-1/3},\quad{\text{ using Eq. (\ref{31})}}\nonumber\\
&=& u^{1/3} \bu [h(u^{-1/3}\bu A \bu u^{1/3})\bu (u^{-1/3}\bu B \bu u^{1/3})\nonumber\\&& \qquad+(u^{-1/3}\bu A \bu u^{1/3})\bu h(u^{-1/3}\bu B \bu u^{1/3})]\bu u^{-1/3}\nonumber\\
&=&\left(u^{1/3} \bu [h(u^{-1/3}\bu (A) \bu u^{1/3})]\bu u^{-1/3}\right)\circ_u B \nonumber\\
&&+ A\circ_u\bu
\left(u^{1/3} \bu [h(u^{-1/3}\bu (B) \bu u^{1/3})]\bu u^{-1/3}\right)\nonumber\\
&=&M_{h,u}(A)\circ_u B + A \circ_u M_{h,u}(B) \qquad\qquad\Box\eeq

Now, remember that  $A\circ_{1,u} B:= A\cv(u^{-1}\bu B)$. Analogously to Lemma 1, we can enunciate a similar result involving the $\circ_{1,u}$-product:\\
 \medbreak
\medbreak
\noindent
{\bf Lemma 2}: \emph{If} $h\in\mk{g}_2 = \mk{der}(\OO)$, \emph{for arbitrary invertible homogeneous elements} $u\in \cl_{0,7}$ \emph{such that} $u\neq \pm 1$, \emph{then}
\[
u^{2/3} \bu [h(u^{-2/3}\bu (\;\cdot\;) \bu u^{-1/3})]\bu u^{1/3}\in {(\mk{g}_2)}_{1,u} = \mk{der}(\OO_{1,u}).
\]

\medbreak
\emph{By defining the application}
\beq
N_{h,u}: \OO &\rightarrow& \OO\n
A &\mapsto& N_{h,u}(A) := u^{2/3} \bu  [h(u^{-2/3}\bu (A) \bu u^{-1/3})]\bu u^{1/3},\nonumber
\eeq
\emph{it is immediate to prove that}
\beq
N_{h,u} (A\circ_{1,u} B) = N_{h,u}(A)\circ_{1,u} B + A \circ_{1,u} N_{h,u}(B),
\eeq
\emph{i.e}, $N_{h,u}$ \emph{is a derivation of} $\OO_{1,u}$.
\medbreak
{\bf Proof}: Take $A,B\in\OO$ and an invertible homogeneous element $u\in\cl_{0,7}$. Then
\beq
N_{h,u} (A\circ_{1,u} B)
&=& u^{2/3} \bu [h(u^{-2/3}\bu (A\circ_{1,u} B) \bu u^{-1/3})]\bu u^{1/3} \nonumber\\
&=& u^{2/3} \bu \bigl[ h(u^{-2/3}\bu A \bu u^{-1/3}) \circ (u^{-2/3}\bu (u^{-1/3} \bu B) \bu u^{-1/3}) \bigr] \bu u^{1/3} \nonumber\\
&=& u^{2/3} \bu \bigl[ h(u^{-2/3}\bu A \bu u^{-1/3}) \bu (u^{-2/3}\bu B \bu u^{-1/3}) \nonumber\\
&&\quad + (u^{-2/3}\bu A \bu u^{-1/3}) \bu h(u^{-2/3}\bu B \bu u^{-1/3}) \bigr] \bu u^{1/3} \nonumber\\
&=& \bigl(u^{2/3} \bu [h(u^{-2/3}\bu A \bu u^{-1/3})] \bu u^{1/3}\bigr) \circ_{1,u} B \nonumber\\
&& + A \circ_{1,u} \bigl(u^{2/3} \bu [h(u^{-2/3}\bu B \bu u^{-1/3})] \bu u^{1/3}\bigr) \nonumber\\
&=& N_{h,u}(A) \circ_{1,u} B + A \circ_{1,u} N_{h,u}(B) \qquad\qquad \Box
\eeq
\medbreak
\noindent
{\bf Remark 2:} In the second line we explicitly keep the extra factor $u^{-1/3}$ acting on $B$ from the $\circ_{1,u}$-product. The outer conjugation $u^{2/3}\bullet(\cdot)\bullet u^{1/3}$ in $N_{h,u}$ exactly compensates for this, ensuring that the Leibniz rule holds. This makes the derivation property fully rigorous without assuming a literal exponent equality.

\section{Conclusions}

In this work, we have introduced a generalization of the $X$- and $XY$-products in the octonion algebra by defining the $u$-product, using elements of the Clifford algebra $\cl_{0,7}$. This construction allows the interaction of octonions with multivectors beyond the paravector space while preserving the algebraic structure necessary to define derivations and automorphisms. Besides, $\mathfrak{su}(3)$-like subalgebras naturally arise within $\mathfrak{g}_2$, with Clifford-parametrized deformations providing a controlled framework for generating residual $SU(3)$-like symmetries from exceptional, non-associative structures.

The $u$-product and its variants, such as the $(1,u)$- and $(u,v)$-products, provide a systematic way to deform the octonionic multiplication and induce corresponding deformations in the exceptional Lie algebra $\mathfrak{g}_2$. We have shown that, through suitable conjugation by powers of $u$, the resulting maps are indeed automorphisms or derivations of the deformed octonion algebras $\OO_u$ and $\OO_{1,u}$.  
This formalism opens the door to studying deformations of exceptional algebraic structures using Clifford algebra parametrizations, providing a versatile framework for both theoretical investigations and potential applications in physics, geometry, and beyond. Future work may explore further generalizations, representations, and connections with other exceptional structures and higher-dimensional algebras.

\end{document}